\documentclass[sigconf]{acmart}

\usepackage{booktabs} 

\setcopyright{rightsretained}

\acmDOI{10.475/123_4}

\acmISBN{123-4567-24-567/08/06}

\acmConference[DEBS'18]{ACM International Conference on Distributed and Event-Based Systems}{June 2018}{Hamilton, New Zealand}
\acmYear{2018}
\copyrightyear{2018}

\acmArticle{4}
\acmPrice{15.00}

\begin{document}
\title{POSTER: Service Discovery for Hyperledger Fabric}

\author{Yacov Manevich}
\affiliation{%
  \institution{IBM Haifa Research Lab}
  \city{Haifa}
  \state{Israel}
}
\email{yacovm@il.ibm.com}

\author{Artem Barger}
\affiliation{%
  \institution{IBM Haifa Research Lab}
  \city{Haifa}
  \state{Israel}
}
\email{bartem@il.ibm.com}

\author{Yoav Tock}
\affiliation{%
  \institution{IBM Haifa Research Lab}
  \city{Haifa}
  \state{Israel}
}
\email{tock@il.ibm.com}

\begin{abstract}
Hyperledger Fabric (HLF) is a modular and extensible permissioned blockchain platform released to open-source and hosted by the Linux Foundation. The platform's design exhibits principles required by enterprise grade business applications like supply-chains, financial transactions, asset management, food safety, and many more. For that end HLF introduces several innovations, two of which are smart contracts in general purpose languages (\emph{chaincode} in HLF), and flexible endorsement policies, which govern whether a transaction is considered valid.

Typical blockchain applications are comprised of two tiers: the first tier focuses on the modelling of the data schema and embedding of business rules into the blockchain by means of smart contracts (\emph{chaincode}) and endorsment policies; and the second tier uses the SDK (Software Development Kit) provided by HLF to implement client side application logic.

However there is a gap between the two tiers that hinders the rapid adoption of changes in the chaincode and endorsement policies within the client SDK. Currently, the chaincode location and endorsement policies are statically configured into the client SDK. This limits the reliability and availability of the client in the event of changes in the platform, and makes the platform more difficult to use. In this work we address and bridge the gap by describing the design and implementation of \emph{Service Discovery}.

\emph{Service Discovery} provides APIs which allow dynamic discovery of the configuration required for the client SDK to interact with the platform, alleviating the client from the burden of maintaining it. This enables the client to rapidly adapt to changes in the platform, thus significantly improving the reliability of the application layer. It also makes the HLF platform more consumable, simplifying the job of creating blockchain applications.
\end{abstract}

%
%



\maketitle

\section{Introduction}
Blockchain technology is gaining a lot of traction becoming one of the most appealing and intriguing areas of interest for both research communities and industrial parties. The popularity of blockchain technologies stems from its huge potential of developing a wide range of distributed applications, allowing safe collaboration between mutually distrusting parties, without the use of a central trusted authority.

Blockchain could be viewed as an append-only immutable data structure - a distributed \emph{ledger} which maintains transaction records between distrusting parties. The transactions are usually grouped into blocks. Then, every party involved in the blockchain network takes part in a consensus protocol to validate transactions and agree on an order between blocks, consequently building a hash chain over these blocks. This process forms a ledger of ordered transactions and is crucial for consistency and integrity. Each party is responsible maintaining its own copy of the distributed ledger not assuming trust on anyone else. Therefore, blockchain protocols exhibits traits that achieve some properties of Byzantine fault tolerance.

Much of the increasing enthusiasm around Bitcoin~\cite{bitcoin} is attributed to blockchain as a promising technology to run trusted exchanges in the digital world. Bitcoin is operated in public, where anyone can join or leave the blockchain network, and no one is required to specify the real identity. Such blockchain systems are known as public or permission-less blockchains. Public blockchains inherently involve the notion of a native cryptocurrency and are mostly based on the \emph{proof-of-work} consensus protocol to compensate for the lack of identity and open group model.
The \emph{proof-of-work} consensus protocol has several salient disadvantages: (1) a huge computational cost, that manifests in prohibitive power consumption, (2) probabilistic nature of transaction confirmation, leading to large confirmation latency, and (3) low transaction throughput. These factors make public blockchains unsuitable for enterprise grade application.
Therefore, growing interest from industry triggered the development of new blockchain platforms designed for permissioned settings, where the blockchain protocol runs among a set of known, authenticated participants.
This is a natural evolution to address requirements posed by business applications running blockchain among a set of identifiable participants which do not fully trust each other. 

It is possible to embed business rules into a Turing complete programmable transaction logic, to be executed by blockchain in the form of a \emph{Smart Contract}, as introduced by Ethereum~\cite{Ethereum}. The Bitcoin script was a predecessor of this concept allowing the transfer of native crypto-coins (bitcoins) from one owner to another. A smart contract provides an abstraction which resembles the functionality of a \emph{trusted distributed application}, leveraging underlying blockchain facilities to gain security and consistency guaranties. Both bitcoin scripts and Ethereum smart contracts resemble a replicated state machine~\cite{SMR}, a well known technique to build resilient distributed applications.
Many permissioned blockchains use the same paradigm: they order the transactions and then execute them on all peers. This is known as the \emph{order-execute} architecture which leads to intolerance to non-deterministic smart contracts and to sequential execution of transactions which severely limits performance~\cite{fabric}.

Hyperledger Fabric~\cite{fabric} (HLF) is an open source project, released to the Linux Foundation\footnote{www.linuxfoundation.org}. It introduces a new architecture for enterprise grade permissioned blockchain platforms following the novel paradigm of \emph{execute-order-validate} for distributed execution of smart contracts (\emph{chaincode} in HLF). In contrast to the \emph{order-execute} paradigm, in HLF transactions are first \emph{executed} by a \emph{subset} of peers (endorsed). Transactions (with results) are then grouped into blocks and \emph{ordered}, and finally a \emph{validation} phase makes sure that transactions were properly endorsed and are not in conflict with other transactions.
This architecture allows multiple transactions to be executed in parallel by disjoint subsets of peers, increasing throughput, and tolerates non-deterministic chaincode. Invalid transactions are dropped in the validation phase.
The \emph{endorsement policy} is the set of rules that determine which subset of peers should execute a transaction, and what constitutes a valid execution. In a sense, HLF benefits from the combination of two well know approaches for replication, passive and active~\cite{budhiraja1993primary, charron2010replication}.

Blockchain applications are typically comprised of two tiers: the first - called the ``platform tier'' - focuses on the modelling of the data schema and embedding of business rules into the blockchain by means of \emph{chaincode} and \emph{endorsment policies}. The second - called the ``client tier'' - uses the SDK (Software Development Kit) provided by HLF to implement client side application logic.
However there is a gap between the two tiers that hinders the rapid adoption of changes in the platform tier within the client tier. Currently\footnote{www.hyperledger.org}, the chaincode identifier and location as well as endorsement policies are statically configured into the HLF client. That is, the client is statically configured with the addresses of the peers that need to execute and endorse a transaction proposal.
This limits the reliability and availability of the client in the event of changes in the platform: whenever the endorsement policy changes, a peer is added or removed, or the chaincode evolves, the client needs to be reconfigured. Moreover, configuration is complicated and technical, which makes the platform more difficult to use.

In this work we describe the design and implementation of the \emph{Service Discovery} component, which extends the architecture and capabilities of HLF, increasing the availability and resiliency of the client side applications. 
Service Discovery provides APIs that allow the client application to dynamically discover the configuration details of the endorsement policies and chaincode it needs to use. It therefore alleviates the client application developer from the burden of painstakingly reconfiguring the client every time these change.
Service Discovery leverages the membership and gossip capabilities of the HLF replication layer~\cite{gossip} to gather and disseminate the necessary information needed to implement theses APIs.

The rest of the paper describes in brief the internal structure of HLF, outlines endorsement policies, and finally presents the design and implementation of the new service discovery component. 
\section{Background}
Prior to Hyperledger Fabric all blockchain platforms, permissioned or permissionless, followed \emph{order-execute} pattern, i.e. network participants use consensus protocol to order transactions and only once the order is decided, all transactions are executed sequentially. Thus essentially implementing active state machine replication~\cite{SMR}. The \emph{order-execute} approach poses a set of limitations, the fact that transactions have to be executed sequentially effectively leads to throughput degradation, becoming a bottleneck. Additionally an important issue to consider which also suffers from the deficiency of the \emph{order-execute} model, is the possible non-deterministic outcome of the transactions. The active state machine replication technique, implies that transaction results has to be deterministic, simply because execution phase followed after consensus-ordering stage to prevent state "forks". Most of the current blockchains implement domain specific language to overcome problem of non-determinism.

\begin{figure}
  \centering
  \includegraphics[width=\columnwidth]{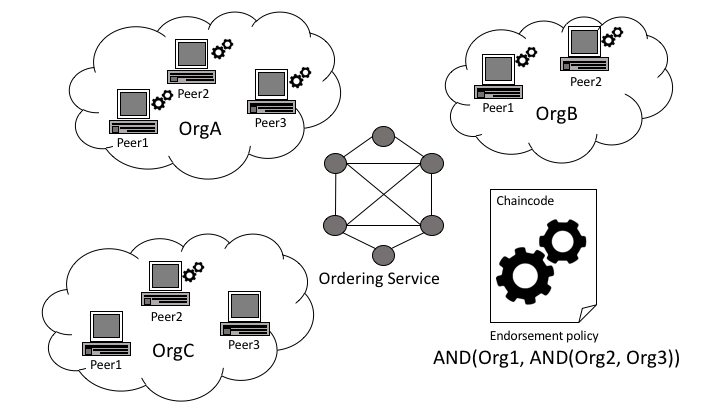}
  \caption{High level structure of Hyperledger Fabric blockchain network. Includes three organizations \textbf{OrgA}, \textbf{OrgA} and \textbf{OrgC}, each including three, two and three peers respectively. The chaincode SampleCC and endorsement policy which requires signature of at least one peer from each organization. And the ordering service which is responsible for total order of transactions.}
  \label{fig:overview}
\end{figure}

Hyperledger Fabric provides modular architecture and introduces a novel \emph{execute-order-validate} approach to address limitations mentioned in the previous paragraph. A distributed application in Hyperledger Fabric basically comprised from two main parts:
\begin{enumerate}
    \item \textbf{Chaincode} - a business logic implemented with general purpose programming language (Java, Go, NodeJS) and invoked during the \emph{execution} phase. The chaincode is a synonym for the well known concept of \emph{smart contracts} and is a core element of Hyperledger Fabric which is executed in a distributed fashion.
    \item \textbf{Endorsement policies} - rules which specify what is the correct set of the peers responsible for the execution and approval of a given chaincode. Such peers, called \emph{endorsing peers}, govern the validity of the chaincode execution results by providing a signature over those results. The endorsement policies are defined with logical expressions such as: $Org1\vee (Org2\wedge Org3)$
\end{enumerate}

The Hyperledger Fabric blockchain network formed by nodes which could be classified into three categories based on theirs roles:
\begin{enumerate}
    \item \textbf{Clients} - network nodes running the application code, which coordinates transaction execation    
    \item \textbf{Peers} - maintain a record of transactions within append-only ledger, responsible for execution of the chaincode and its lifecycle. In order to allow load balancing, not all peers are responsible for execution of the chaincode, but only a subset of peers called \emph{endorsing peers}
    \item \textbf{Ordering nodes} - a cluster of the replica nodes which exposes an abstraction of atomic broadcast to establish total order between all transactions within Hyperledger Fabric. Ordering nodes are completely oblivious to the application state and don't take any part in transaction validation or execution.
\end{enumerate}

In order to provide finer grained privacy and confidentiality Hyperledger Fabric introduces concept of \emph{channels}, a high level abstraction which basically represents a blockchain network. Each channel can contain different or even disjoin set of peers, thus allowing to segregate application state allowing greater privacy control by partitioning data across different channels.

\subsection{Transaction execution flow}\label{tx_execution}
\begin{figure}
  \centering
  \includegraphics[scale=0.25]{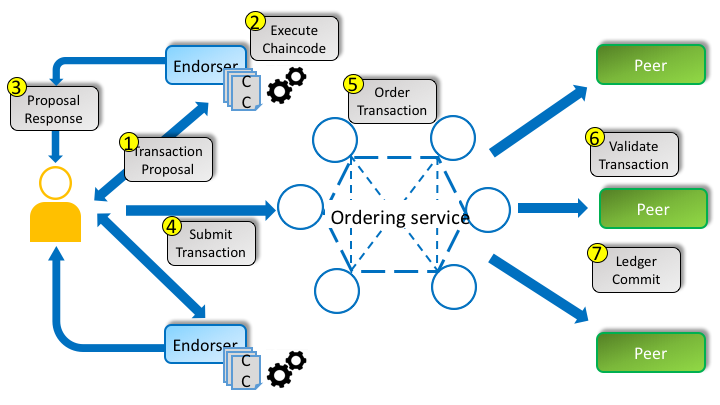}
  \caption{Hyperledger Fabric - high level transaction flow.}
  \label{fig:flow}
\end{figure} 
The following summarises the execution flow of transaction submitted by a client into Hyperledger Fabric, depicted in Fig.~\ref{fig:flow}:

\begin{enumerate}
    \item Client uses SDK to form a \emph{transaction proposal}, which includes: the channel name, the chaincode name to invoke and the input parameters for the chaincode to be executed. Next, client sends transaction proposal to all endorsing peers to satisfy the endorsement policy of the given chaincode.
    \item Endorsing peers simulate the transaction based on parameters received from the client, by actually interacting with chaincode  to record state updates and produce output in the form of read-write set, following by signing the read-write set and returning the results back to the client.
    \item Client collects responses from all endorsing peers, validates that results are consistent, e.g. all endorsing peers have signed the same payload, followed by concatenation of all signatures of the endorsing peers along with the read-write sets, creating a transaction which is submitted to the ordering service.
    \item Ordering service collects all incoming transactions, order them to impose total order of transactions within channel context and periodically cuts blocks which include all those transactions ordered. 
    \item Dedicated peers of each organization, pull new blocks from the ordering service and disseminate then by using scalable middleware for ledger replication, which implementation is based on an epidemic diffusion based protocol - gossip~\cite{gossip}.
    \item Each peer upon receiving a new block, iterates over transactions to validate: a) the endorsement policy, i.e. whether the set of the endorsing peers signatures satisfies the endorsement policy correlated to the chaincode; b) performs multi-value concurrency control checks.
    \item Once the transaction validation has finished, the peer appends the block to the ledger and updates its state based on valid transactions. After the block is committed it emits events to update the client connected to it.
\end{enumerate}

\section{Service Discovery}

In order to execute chaincode on peers, submit transactions to orderers, and to
be updated about the status of transactions, applications connect to an API
exposed by an SDK as outlined in section~\ref{tx_execution}.

However, the SDK needs a lot of information in order to allow applications to
connect to the relevant network nodes. In addition to the enrollement CA and TLS CA certificates
of the orderers and peers on the channel - as well as their IP addresses and port
numbers - it must know the relevant endorsement policies along with which peers
have the chaincode installed (so the application knows which peers to send chaincode
proposals to) on them.

In previous versions of Hyperledger Fabric, this information was statically encoded. 
However, this implementation is not dynamically reactive to network changes (such as the 
addition of peers who have installed the relevant chaincode, or peers that are temporarily offline). 
Static configurations also do not allow applications to react to changes of the
endorsement policy itself (as might happen when a new organization joins a channel).

Furthermore, the client application has no way of knowing which peers have updated ledgers
and which do not, so it might submit proposals to peers whose ledger data is not in
sync with the rest of the network, resulting in transaction being invalidated upon
commit. This is a waste of both time and resources.

The \emph{discovery service} improves this process by having the peers compute
the needed information dynamically and present it to the SDK in a consumable
manner.

\subsection{How service discovery works in Fabric}

The application is bootstrapped knowing about a group of peers which are
trusted by the application developer/administrator to provide authentic responses
to discovery queries. A good candidate peer to be used by the client application
is one that is in the same organization.

The application issues a configuration query to the discovery service and obtains
all the static information it would have otherwise needed to communicate with the
rest of the nodes of the network. This information can be refreshed at any point
by sending a subsequent query to the discovery service of a peer.

The service runs on peers -- not on the application -- and uses the network metadata
information maintained by the gossip~\cite{gossip} communication layer to render the list of peers
that are online. It also fetches information, such as relevant endorsement policies,
from the peer's state database.

With service discovery, applications no longer need to specify which peers they
need endorsements from. The SDK can simply send a query to the discovery service
asking which peers are needed given a channel and a chaincode ID. 

The discovery service can respond to the following queries:

\begin{itemize}
    \item \textbf{Configuration query} - returns the configuration required for initialization of the CA certificates of all organizations in the channel along with the orderer endpoints of the channel.
    \item \textbf{Peer membership query} - returns the peers that have joined the channel.
    \item \textbf{Endorsement query} returns an endorsement descriptor for given chaincode(s). The descriptor allows easy selection of some set of peers such that if endorsements are obtained from the set, the endorsement policy would be satisfied.
    \item  \textbf{Local peer membership query} returns the local membership information of the peer that responds to the query.
\end{itemize}

\bibliographystyle{ACM-Reference-Format}
\bibliography{sample-bibliography}

\end{document}